# Single photon interference between two modes originated from a single quantum dot

H. Kumano[1,2], S. Ekuni[1], H. Nakajima[1], M. Jo[1], H. Sasakura[1], S. Adachi[3], S. Muto[3], and I. Suemune[1,2]

[1]*RIES, Hokkaido University, Sapporo 001-0021, Japan*
[2]*Japan Science and Technology Corporation (CREST), Saitama 332-0012, Japan 332-0012, Japan*
[3]*Graduate School of Engineering, Hokkaido University, Sapporo 060-8628, Japan*

Interference of a single photon generated from a single quantum dot is observed between two photon polarization modes. Each emitted single photon has two orthogonal polarization modes associated with the solid-state single photon source, in which two non-degenerate neutral exciton states are involved. The interference between the two modes takes place only under the condition that the emitted photon is free from which-mode information.

Young's double slit experiments [1-6] have successfully revealed one of the most fundamental aspects of quantum mechanics; Wave-particle duality as a manifestation of the quantum complementarity [7, 8] for quantum mechanical object such as matter waves and photons. So far, the double slit experiments have been demonstrated for massive particles such as neutrons [1], atoms[2], single electrons[3] and single molecules[4], and also for the massless photon with attenuated coherent light[5] and with single photons from a color center in a diamond nanocrystal [6]. In most of these Young's double slit type experiments, two modes to generate interference are separately prepared from each sources, and so far the interference between modes formed *in* solid-state systems has only been demonstrated for electrons in semiconductors [9,10].

In this paper, single photon interference between two photon polarization modes, which stems from non-degenerate neutral exciton states in a single quantum dot (QD) is demonstrated. Interference between the two polarization modes appears only when one can not recognize "which-mode" a emitted single photon is in. The duality of the present experiments with the Young's double-slit experiments exists, i.e., the photon polarization modes instead of the spatial modes as well as the detection polarizations angles rather than the position on a screen. It is noted however that the present inference experiments take place in a single QD.

An $In_{0.75}Al_{0.25}As$ QDs sample was grown on a semi-insulating (001)-GaAs substrate by molecular-beam epitaxy. The QDs were prepared in Stranski-Krastanow (S-K) growth mode on $Al_{0.3}Ga_{0.7}As$ layers and were sandwiched with $Al_{0.3}Ga_{0.7}As$ layers. The topmost surface was terminated with a GaAs cap layer. After the growth, the sample was etched into mesa structures with diameters of ~150 nm for isolating single QD from the dot ensemble with the density of around $5 \times 10^{10}$ dots/cm$^2$. Further details of this sample preparation are described in Ref. 11. For a single dot spectroscopy, a continuous-wave Ti: sapphire laser was used as a linearly polarized excitation source. An objective lens with the numerical aperture of 0.42 focused the laser beam on one of the mesa structures and collected photoluminescence (PL) emitted from the mesa.

In order to discuss single photon interference between two polarization modes, luminescence collected under non-resonant excitation (1.687 eV) was analyzed by a half wave plate (HWP) followed by a Glan-Thompson polarizer (GTP) placed in front of two separate detection systems; (i) a 0.64-m triple-grating spectrometer equipped with a liquid-nitrogen cooled Si charge-coupled-device (CCD) detector (energy resolution of this detection system < 5 µeV) and (ii) single photon counting module (EG&G SPCM) with a filter to select specific emission line as depicted in Fig. 1 (a). Transmission axis of the GTP is set vertical in the laboratory frame and aligned with the crystalline axis of the sample. Two quantum dots (QD A, QD B) with different exciton fine structure splitting (FSS) were studied at 20 K. Generation of single photons from both QDs has previously been confirmed independently by the photon correlation measurements [12-14]. Throughout this work, measurements were carried out under the weak excitation condition below 50 µW, so that the second-order photon correlation function at zero time delay $g^{(2)}(0)$ would not show noticeable degradation indicating multi-photon emission [15]. Therefore only single photon regime is responsible for the results discussed hereafter.

In QD A, four emission lines located at 1.5976 eV (neutral exciton: $X^0$), 1.5964 eV (neutral bi-exciton: $XX^0$), 1.5966 eV (negatively charged exciton: $X^-$), and 1.6009 eV (positively charged exciton: $X^+$) were dominant and the origin of each line was identified with several kinds of experiments [15]. In this paper, in order to discuss the impact of which-mode information on a single photon interference in a rectilinear polarization basis, we focus on the neutral exciton emission. Linear-polarization dependence of emitted photons recorded by the CCD detector is displayed as a contour plot in Fig. 1(b). In this figure the horizontal axis indicates the polarization rotation angle $2\theta$, ($\theta$ is defined as a fast axis angle of the HWP measured from the vertical), while the vertical axis shows the photon energy. Hereafter, the angle dependences will be discussed with $2\theta$. It is found that emitted photons are linearly polarized, which constitutes photon polarization modes, V and H. The transition energy of the $X^0$ line showed the corresponding FSS. Its splitting energy of 110 µeV originates from the anisotropy of the confinement potential [16] as depicted in Fig. 1(c).

Figure 2(a) shows the PL spectra measured with vertical (V) and horizontal (H) polarizations and the solid lines are fitting with the Lorentzian function. The FSS of 110 µeV exceeds the linewidth (full width at half maximum) of 78 µeV, and therefore the two spectra are well separated in energy and are associated with the two polarization modes. The integrated PL intensity over the whole spectral region of the $X^0$ emission is shown by the open circles in the upper panel of Fig. 2(b). It remains constant against the detection polarization angle. For comparison, detection polarization dependence of the $X^0$ transition energy (blue crosses) is also shown in the lower panel of Fig. 2(b) together with that of $XX^0$ (gray crosses). Here the vertical axis indicates the relative peak energy shifts measured from the respective mean transition energies of $X^0$ and $XX^0$ displayed against the detection polarizations. The clear correlation between the emission energy and polarization is observed, which gives us the polarization angle reference.

Figure 2(c) shows PL spectra of QD B measured with V and H polarizations and their Lorentzian fit (solid curves). In the QD B, FSS and the linewidths are 30 μeV and 100 μeV, respectively. This gives the large spectral overlap and make the transition energy labeling to the two polarization modes indefinite, leading to less which-mode information in this QD B. A series of analyses similar to those of the QD A were also carried out with the QD B, and the results are summarized in Fig. 2(d). In striking contrast to the QD A, the integrated PL intensity exhibits clear periodic change depending on the detection polarizations. In addition, π/4 phase shift from the vertical is obvious, where the intensity is maximized at π/4+nπ (*n*: integer) denoted as D and minimized at 3π/4+nπ denoted as D*.

With the same amplitude for both polarization modes ensured by Fig. 2(a) and (c), the single photon state generated by the radiative recombination of a neutral exciton in a single QD can be described, for two extreme cases, as (i) $\rho = 1/2(|V\rangle\langle V| + |H\rangle\langle H|)$ for mixed states, and (ii) $\rho = |\psi\rangle\langle\psi|$ for pure state $|\psi\rangle = \frac{1}{\sqrt{2}}(|V\rangle + |H\rangle)$, which is a superposition of two polarization modes. Jones matrix P(π) for the HWP with its fast axis in the vertical direction is given by $\begin{pmatrix} 1 & 0 \\ 0 & e^{i\pi} \end{pmatrix}$ under $|V\rangle = (1,0)^T$ and $|H\rangle = (0,1)^T$ basis [17]. Since the GTP is described by a projection operator $\Lambda_V = |V\rangle\langle V|$, photon field operator E(θ) via these HWP and GTP is given by $\Lambda_V \cdot R(\theta)P(\pi)R(\theta)^{-1}$, where R(θ) is a rotation matrix expressed by $\begin{pmatrix} \cos\theta & -\sin\theta \\ \sin\theta & \cos\theta \end{pmatrix}$. In the mixed photon state, integrated PL intensity I(θ) given by $\text{Tr}\{\rho E(\theta)^\dagger E(\theta)\}$ leads to the constant value of 1/2 independent of θ, which agrees well with the experimental result for the QD A (Fig. 2(b)). On the other hand, for the 1 photon 2 mode state (pure state) I(θ) becomes $\frac{1}{2}[1 + \cos 2(2\theta - \pi/4)]$ (1), which reproduces the experimental polarization angle dependence of the QD B including modulation period of π and phase shift of π/4 as shown in the solid curve in the upper panel of Fig. 2(d).

With respect to the above argument based on the experiments with the high-resolution triple-grating followed by multi-channel detector, information on the polarization mode of a detected photon is present if linewidth < FSS since the energy resolution of the detection system is high enough. In this case, the photon state is reduced to the mixed state and the interference disappears. On the other hand, if linewidth > FSS, spectral overlap measured under the two polarizations are large enough not to label the two polarization modes by their transition energies, which forms a superposition state leading to the two-mode interference. It is worth noting that which-mode information, *i.e.*, possibility to discriminate which mode the emitted photon belongs to, is given by a grating which can spatially separate photons with energy difference comparable to the FSS. Thus when the photon is not dispersed by the grating in a sense that the mode discrimination is unattained due to the low energy resolution, the polarization modes will never be labeled by their transition energies and the which-mode information is absent even in a QD with linewidth < FSS. Along this idea, photon intensity as a function of the polarization angle $2\theta$ is examined with a SPCM after the $X^0$ line filtering by a weakly-diffractive grating [18] as depicted in Fig. 1(a) following dotted arrows. In this measurement, QD A was employed because of the presence of which-mode information. Resultant intensity behavior without energy labeling to the modes is shown in Fig. 3, and the interference clearly emerges as expected. The observed π phase shift in comparison to the Fig. 2(d) originates from the opposite elongation direction [19] which exchanges the mode definition.

The observed visibilities for the two QDs in Fig. 2(d) and Fig. 3 are 0.09 (QD A) and 0.12 (QD B), respectively. These values are rather low relative to the unity expected for the case of perfect interference (Eq. 1). In general, inequality $\mathcal{P}^2 + \mathcal{V}^2 \leq 1$ holds based on the wave-particle duality, where $\mathcal{P}$ is a predictability of the which-mode (indicating particle-like nature), and $\mathcal{V}$ is a visibility (wave-like nature) [20]. The equality occurs only in the absence of any incoherent contribution. For the SPCM detection without mode labeling, predictability of the which-mode can be safely assumed to be zero. Therefore, it is probable that the incoherent contribution is dominant for the degraded visibility in the present QDs. Contribution of the relative phase between the modes which we have ignored in this study could be another origin.

In summary, two-mode interference of single photon is observed with non-degenerate neutral exciton states in a single quantum dot. Since the neutral exciton states define the single photon polarization modes, the energy relaxation process in solid state system serves as a mode distributor. The one-photon two-mode interference takes place only under the condition that the polarization modes of emitted photon is indistinguishable in energy, thus the which-mode information is absent.

The authors would like to acknowledge Dr. H. Z. Song, S. Hirose and M. Takatsu for the sample preparation, and R. Kaji for fruitful discussions. This work was supported in part by the Grant-in-Aid for Scientific Research (A)(2), No. 21246048, Young Scientists (A), No. 2168102009, and Hokkaido Innovation Through Nanotechnology Supports (HINTs) from the Ministry of Education, Culture, Sports, Science and Technology.

Figure captions

Fig. 1

(color on line) (a) Schematic of the experimental setup. Two detection systems are used; (i) a triple-grating and subsequent multi-channel detector and (ii) a single photon counting module (SPCM) connected with a filter. (b) Contour plot of detection energy as a function of polarization rotation angle $2\theta$ for the $X^0$ in a QD A. Detection polarization directions are illustrated. (c) Energy structure and fine structure splitting (FSS) of the $X^0$ states in a QD with anisotropic potential.

Fig. 2

(color on line) (a) Emission lineshape of the $X^0$ in QD A for both polarization modes $|V\rangle$ and $|H\rangle$. (b) (lower panel) Detection polarization dependence of the relative $X^0$ (blue crosses) and $XX^0$ (gray crosses) peak energy shifts measured from the mean value. (upper panel) Normalized PL intensity of the $X^0$ integrated over both polarization modes. (c) Emission lineshape of the $X^0$ in QD B. Additional peak at ~1.59395 eV is $X^+$ line. (d) Results of the same experiments as (b) for the QD B. Here the fitted intensity with $\frac{1}{2}[1+\mathcal{V}\cos 2(2\theta+\theta_0)]$ is also shown as a solid curve. Resultant visibility $\mathcal{V}$ is 0.12.

Fig. 3

(color on line) Normalized intensity detected with a SPCM as a function of $2\theta$ for the $X^0$ in a QD A. Solid curve is a fitted result with $\frac{1}{2}[1+\mathcal{V}\cos 2(2\theta+\theta_0)]$, where the resultant $\mathcal{V}$ is 0.09.

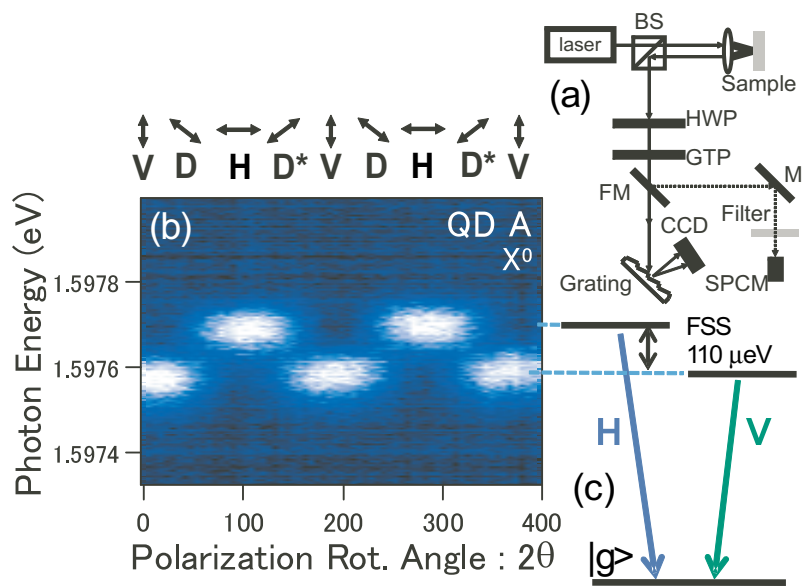

H. Kumano et al., Figure 1

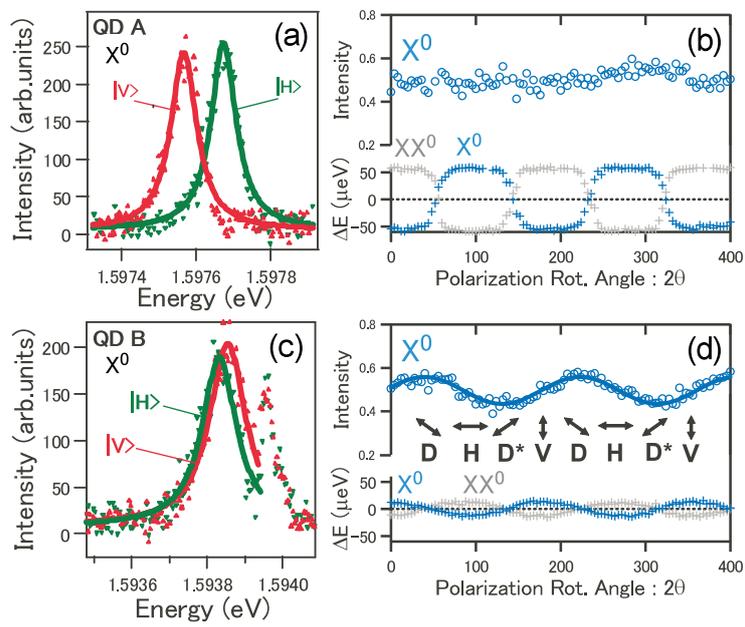

H. Kumano et al., Figure 2

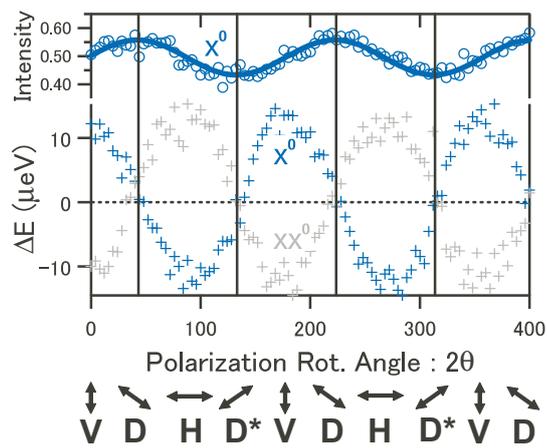

H. Kumano et al., Figure 3